\documentclass{czjphys}         
\usepackage{amssymb,cite,graphicx,amsmath}
\newcommand{\ri}{\right}
\newcommand{\lf}{\left}
\newcommand{\CaC}{{\cal C}}

\newcommand{\PT}{{\cal P}{\cal T}}
\newcommand{\fract}[2]{{\textstyle\frac{#1}{#2}}}

\newcommand\ZZ{{\mathbb Z}}
\newcommand\TT{{\mathbb T}}

\newcommand\LL{{\mathbb L}}

\newcommand{\bbalpha}{\boldsymbol{\alpha}}
\newcommand{\Dd}{{\rm D}}
\begin{document}
\title{A reality proof in $\PT$-symmetric quantum mechanics}
%
\authori{Patrick Dorey}     \addressi{Department of
  Mathematical Sciences,
         University of Durham,
         Durham DH1 3LE,
         UK.}
\authorii{Clare Dunning}      \addressii{Department of Mathematics,
  University of York, York, YO1 5DD, UK.\footnote{Address after
  September 2003: Department of Mathematics, University
 of Queensland, Brisbane, QLD 4072, Australia.}
}
\authoriii{Roberto Tateo}
\addressiii{Department of
  Mathematical Sciences,
         University of Durham,
         Durham DH1 3LE,
         UK.\footnote{{\rm
        \uppercase{A}ddress after November 2003: \uppercase{D}ipartimento di
            \uppercase{F}isica \uppercase{T}eorica,
         \uppercase{U}niversit\`a di \uppercase{T}orino,
         \uppercase{V}ia \uppercase{P}.~\uppercase{G}iuria 1, 10125
         \uppercase{T}orino, \uppercase{I}taly.}}
}
\authoriv{}     \addressiv{}
\authorv{}      \addressv{}
\authorvi{}     \addressvi{}
%
\headauthor{ Dorey et al.}            
\headtitle{ \ldots}             
\lastevenhead{ et al.:  \ldots} 
\refnum{A}
\daterec{12/09/03}    
\issuenumber{0}  \year{2003}
\setcounter{page}{1}
\maketitle

\begin{abstract}
We review the proof of a conjecture concerning the reality of the
 spectra of certain $\PT$-symmetric quantum mechanical systems,
 obtained via a connection between the theories of
 ordinary differential equations and integrable models.
Spectral equivalences inspired by the correspondence are also discussed.
\end{abstract}

\section{Introduction}
The unexpected reality of the spectra of a class of non-Hermitian quantum
mechanical systems has provoked a fair amount of work since the first
example was noted by Bessis and Zinn-Justin \cite{BZJ}.
Generalising the result at $M=3/2$, a numerical
 study of  the Schr\"odinger equations
\begin{equation}
-\frac{d^2}{dx^2}\psi(x)-(ix)^{2M}\psi(x)=E\psi(x)
\label{bbeq}
\end{equation}
led Bender and Boettcher to conjecture that under suitable boundary
conditions the spectrum of (\ref{bbeq}) was entirely real and positive
 provided $M \ge 1$ \cite{BB}.

Bender and Boettcher also noticed the
relevance of $\PT$-symmetry \cite{BB,BBN} to these problems.
For all $M$, the  potentials $V(x)=-(ix)^{2M}$
are invariant
under the combined action of parity and
time reversal, and this implies
 the eigenvalues are either real or occur in complex-conjugate pairs.
If all energy eigenstates  are
also eigenstates of the $\PT$ operator,
then the spectrum is guaranteed to be entirely real,
but this is not true in general:
 the $\PT$-symmetry can be
 spontaneously broken, in which case the corresponding energies occur
in complex-conjugate pairs.
In fact, a proof of the reality conjecture
has only recently been obtained \cite{DDTb}, via a correspondence
between
ordinary differential equations and integrable models
(the
so-called `ODE/IM
correspondence' \cite{DTa,BLZa,DTb,Sc}).  Here we review the proof and
touch on a few of the byproducts of the correspondence relevant to
the ODE community.

Generalising the Bessis -- Zinn-Justin and Bender -- Boettcher
Hamiltonians, we consider the class of
 $\PT$-symmetric  problems
\begin{equation}
\lf [ -\frac{d^2}{dx^2} -(ix)^{2M} - \alpha
(ix)^{M-1}+\frac{l(l+1)}{x^2} \ri] \psi(x) = \lambda
\psi(x) \quad,\quad \psi(x) \in L^2(\CaC)~.
\label{PTg}
\end{equation}
For $M<2$,
the contour $\CaC$ may be taken to be the real axis with a small
dip below the origin if $l(l+1) \neq 0$. For larger $M$,
the contour should be
distorted into the complex
plane to ensure the correct analytical continuation of the
original problem,
as explained in \cite{BB}.
The result of \cite{DDTb}
 is that the spectrum of (\ref{PTg}) is
{\it real} for $M\ge 1$ and $\alpha <   M+1+|2l+1| $
and
{\it positive} for $M\ge 1$ and $\alpha < M+1-|2l+1|$~.
Referring to figure~\ref{regions}, reality was proven for
 $(\alpha,l)\in B\cup C\cup D$, and positivity for
$(\alpha,l)\in D$.
\bfg[t]                     
\bc                         
\includegraphics[width=0.42\linewidth]{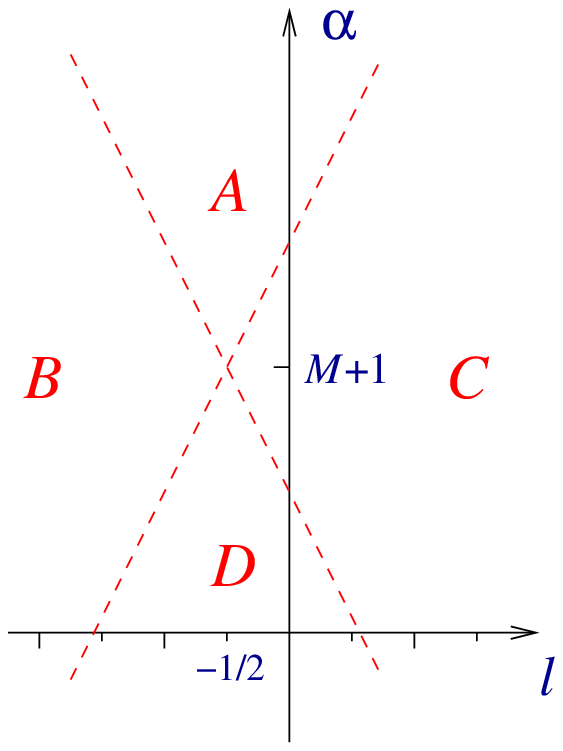}
\ec                         
\vspace{-12mm}
\caption{\label{regions} The initial `phase diagram' at fixed $M$.
}
\efg                        
\section{The functional relation}
Rather than considering the $\PT$-symmetric problem  (\ref{PTg})
alone,  we introduce a
related problem obtained from (\ref{PTg})
by sending $x \to x/i$ and $E \to -E$:
\begin{equation}
\left(-\frac{d^2}{dx^2}+x^{2M}\pm\alpha
  x^{M-1}+\frac{l(l{+}1)}{x^2}\right)\Phi(x)
= E
\Phi(x)~.
\label{eqd}
\end{equation}
It will be convenient to treat positive and negative values of
$\alpha$ together.
The boundary conditions for (\ref{eqd})
are that $\Phi(x)$ should vanish as $x \to \infty$ along the
real axis, and
behave as $x^{l+1}$ as $x \to 0$.   For $l$ real and larger than $-1/2$,
this problem is
Hermitian. In the language of ordinary differential equations in the
complex domain, it is sometimes called a `radial' problem, while
(\ref{PTg}) is called a `lateral' problem.
(Note, the radial problem can also be considered for $\Re e\,l\le
-1/2$,
but is best then defined by analytic continuation in $l$.)

Following the ideas particularly advocated by
Sibuya and Voros  in the context of ordinary
differential equations \cite{Sha,Voros}, we
approach the spectral problems by
studying the behaviour  of the  associated spectral
determinants. Adopting the convention used in \cite{DDTa}, let
 $\{E_j^{(\pm)}\}$ be the set of eigenvalues of
(\ref{PTg}) with inhomogeneous term $\mp\alpha  (ix)^{M{-}1}$, and let
$\{e_i^{(\pm)}\}$ be the
eigenvalues of (\ref{eqd})  with inhomogeneous term $\pm\alpha x^{M{-}1}$.
Then define
two pairs of spectral determinants, as follows:
\begin{equation}
T^{(\pm)}(E)
=T^{(\pm)}(0)\prod_{j=0}^{\infty}\left(1+\frac{E}{E^{(\pm)}_j}\right)~,
\label{specdetT}
\end{equation}
and
\begin{equation}
Q^{(\pm)}(E) =Q^{(\pm)}(0)\prod_{i=0}^{\infty}\left(1-\frac{E}{e^{(\pm)}_i}\right)~.
\label{specdeta}
\end{equation}
Both products are convergent for $M>1$, and define
entire functions of $E$, and their zeroes (or, for $T(E)$, their negatives)
coincide with the eigenvalues of the
corresponding spectral problem. For $M\le 1$
convergence factors must be added, and it is more efficient
to define the spectral determinants indirectly, via certain special
`Sibuya' solutions to (\ref{eqd}), which are anyway needed to prove
the key identity (\ref{tqeqb}) below; see \cite{Sha,DTb} for more details.

By considering the asymptotic behaviour of the Sibuya solutions in
the complex plane it is
not hard, following the arguments given in \cite{DTb},
to establish a Stokes relation, from which
one can obtain the following functional equations
\begin{equation}
T^{(\pm)}(E)Q^{(\pm)}(E)= \omega^{-(2l+1\pm\alpha)/2} Q^{(\mp)}(\omega^{-2}E) +
 \omega^{(2l+1\pm\alpha)/2} Q^{(\mp)}(\omega^{2}E)~,
\label{tqeqb}
\end{equation}
where $\omega= e^{i\pi/(M+1)}$. This shows that the spectral problems
(\ref{PTg}) and (\ref{eqd}) are related by much more than a simple
change of variables and boundary conditions, and also that the spectral
problem at positive $\alpha$ is necessarily tied up with the
problem at negative $\alpha$.
Furthermore, this
functional relation taken at $\alpha=0$ is well-known in the
integrable model (IM) world, where
it goes by the name of
Baxter's TQ relation.
 This observation, initially developed in \cite{DTa,BLZa,DTb}, has led
 to an  exact mapping
between ODE quantities such as spectral determinants and
functions constructed  in the context of
integrable models. It provides a powerful tool to analyse ODE problems
using IM techniques, and vice-versa.

If we set $E=e^{(\mp)}_j$ in (\ref{tqeqb}) we can use the entirety of $T$
and the product form for $Q$ (\ref{specdeta}) to obtain
\begin{equation}
\prod_{i=0}^{\infty}\left(\frac{e_i^{(\mp)}-
\omega^2e^{(\pm)}_j}{e^{(\mp)}_i-\omega^{-2}e^{(\pm)}_j}\right)
 = -\omega^{-2l-1\mp\alpha}\,.
\label{bae1}
\end{equation}
In IM language sets of coupled equations of this type are known as
Bethe ansatz equations. Their solution allows $Q$ to be reconstructed
  via (\ref{specdeta}) and thus $T$ using the TQ relation
(\ref{tqeqb}).  Finally the  eigenvalues of the $\PT$-symmetric
Hamiltonian are found by searching for the zeros of $T(E)$.
Alternatively,  a technique developed in
integrable models allows one to
convert the Bethe ansatz equations (\ref{bae1}) into a single
nonlinear integral equation, from
which one can easily numerically obtain both the zeros of $Q(E)$ and  $T(E)$,
as first emphasised in \cite{DTa,DTb}.

\section{The proof}
Returning to the reality proof, a second set of Bethe ansatz like
equations can be obtained from
(\ref{tqeqb}) if we instead set  $E=-E^{(+)}_j$ and invoke the entirety of
 $Q^{(-)}(E)$.
Rearranging and using the product form (\ref{specdeta})  leads to
\begin{equation}
\prod_{i=0}^{\infty}\left(\frac{e^{(-)}_i+\omega^2E^{(+)}_j}
{e^{(-)}_i+\omega^{-2}E^{(+)}_j}\right)
 = -\omega^{-2l-1-\alpha}\,.
\label{bae2}
\end{equation}
This equation couples the so-far mysterious eigenvalue
$E^{(+)}_j$
of the $\PT$-symmetric problem (\ref{PTg}) to the much
better-controlled  eigenvalues of
the Hermitian problem (\ref{eqd}).  Indeed,
a Langer transformation \cite{La} shows that the
 eigenvalues $e^{(-)}$
of the radial problem are also positive if $\alpha < M+2l+2$
\cite{DDTb}. It is from this equation that we
are able to prove the reality of the $\PT$-symmetric problem.

The
first step will be to
set $E^{(+)}_j = |E^{(+)}_j| \exp(i \,\delta_k)$.  Now take the modulus$^2$
 of (\ref{bae2}) to obtain
\begin{equation} 
\prod_{i=0}^{\infty} \lf ( \frac{ (e^{(-)}_i)^2 + |E^{(+)}_j|^2  + 2
  \, e^{(-)}_i \, 
 |E^{(+)}_j| \cos(\fract{2 \pi}{M+1} + \delta_j)}
{
(e^{(-)}_i)^2 + |E^{(+)}_j|^2  + 2  \, e^{(-)}_i\,
 |E^{(+)}_j| \cos(\fract{2\pi}{M+1} - \delta_j)} 
\ri)
= 1\, .
\label{abs}
\end{equation}
For $\alpha<M+2l+2$ all the $e^{(-)}_i$ are positive,
and each single term in the product on the LHS of (\ref{abs}) is either
greater than, smaller than, or equal to one
depending  only on the relative values of the cosine terms in the numerator
and denominator. These are
independent of the index $i$.
Therefore the only possibility to match the RHS is for each term in the
product to be individually equal to one, which for $E^{(+)}_j\neq 0$
requires
\begin{equation} 
\cos(\fract{2\pi}{M+1}+\delta_j)=\cos(\fract{2\pi}{M+1}-\delta_j) ~,
\quad
\mbox{or}
\quad 
\sin(\fract{2\pi}{M{+}1})\sin(\delta_j)=0\,.
\end{equation}
Since  $M > 1$, this latter condition implies
\begin{equation}
\delta_j= n \pi\, ,~~~~n \in \ZZ
\end{equation}
and this establishes the reality of the
eigenvalues of (\ref{PTg})
 for $M>1$ and $\alpha<M+2l+2$ or, relaxing the condition on
$l$, $\alpha<M+1+|2l{+}1|$.

At $M\le1$
the simple form (\ref{specdeta}) for $D^{(-)}(E)$ acquires extra
 convergence factors, resulting in a break down of the proof,
as expected  given the numerical
findings of \cite{BB,DTb}. More details can
be found in \cite{DDTb}; for a further
generalisation of the method, see \cite{Shin:2002vu}.

Finally, we remark  that while the above constraints on the parameters
$M,\alpha$ and $l$ are
sufficient conditions  they are
not necessary, as demonstrated in  figure~\ref{scan} for the case $M=3$.
The full domain of unreality, obtained numerically in \cite{DDTc}, is
the interior of the curved line,  a proper
subset of $A$ which only touches its boundary at isolated points.
In the small, approximately-triangular region inside $A$ but
outside the
curved line which abuts the  crossing of the lines separating the
regions $A,B,C,D$,
the spectrum
is not only real but also entirely positive, despite the fact that it lies
outside the domain $D$.  The separating lines themselves have further
significance, in that at these points the model has a hidden
supersymmetry, which plays a significant r\^ole in the breakdown of
reality \cite{DDTc}.
\bfg[t]                     
\bc                         
\includegraphics[width=0.42\linewidth]{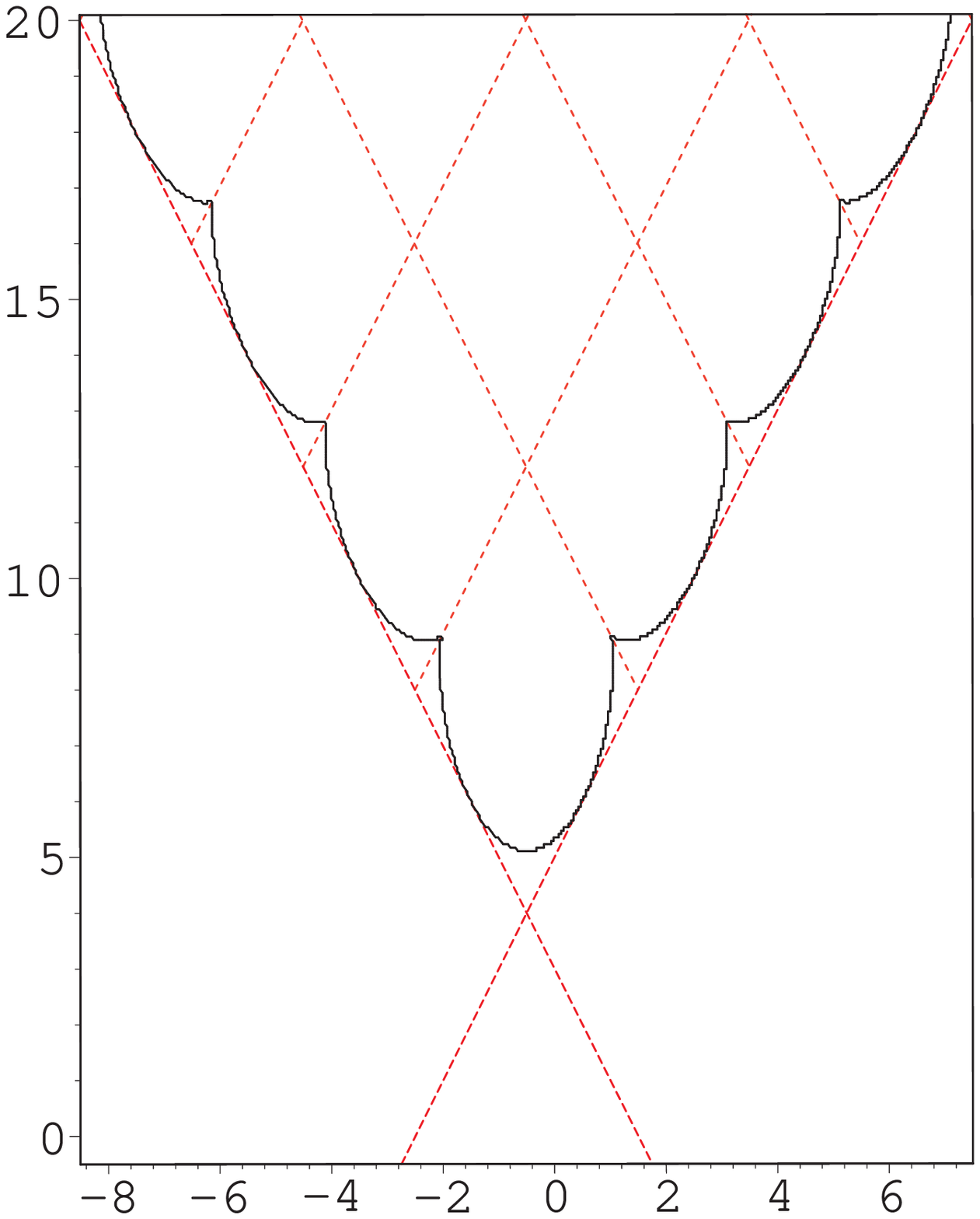}
\ec                         
\vspace{-7mm}
\caption{\label{scan}The domain of unreality for $M=3$. Horizontal and
vertical axes are $l$ and $\alpha$ respectively.}
\efg                        

\section{Further consequences}
One further application of the ODE/IM correspondence was also found in
\cite{DDTb}, and concerns
the lateral and
radial problems at $M=3$.
As remarked in \cite{DDTrevc},
it is convenient to reparametrise the
angular-momentum term by setting $\lambda=\sqrt{3}(2l{+}1)$\,, so that
the equation becomes
\begin{equation}
\left(-\frac{d^2}{dx^2}+x^{6}+\alpha x^2+
\frac{\lambda^2-3}{12\,x^2}\right)\Phi(x) = E
\Phi(x)~.
\label{eqe}
\end{equation}
Define  $\bbalpha=(\alpha,\lambda)^T$ and
let $D(E,\bbalpha)$ denote the spectral determinant for this problem.
Via the Bethe ansatz approach, it turns out that this problem has a
relationship with a {\it third}-order ordinary differential
equation:
\begin{equation}
\left( \Dd(g_2{-}2)\Dd(g_1{-}1)\Dd(g_0) + x^3\right)\phi=
\frac{3\sqrt{3}}{4}\,E\phi
\end{equation}
where $\Dd(g)\equiv (\frac{d}{dx}-\frac{g}{x})$, and
\begin{equation}
g_0=1+(\alpha+\sqrt{3}\lambda)/4~~,~~
g_1=1+\alpha/2~~,~~
g_2=1+(\alpha-\sqrt{3}\lambda)/4~.
\end{equation}
This third-order equation is associated with $SU(3)$ Bethe ansatz
equations, as discussed in \cite{DTc,DDTa}. Furthermore, the
third-order equation is symmetrical in $\{g_0,g_1,g_2\}$, a
feature which is completely hidden in the original second-order
equation. By playing with this symmetry, one can establish some
novel spectral equivalences
 between different
(second-order) radial problems, and also between these and certain
lateral problems.
We add to the results discussed in \cite{DDTb}
the remark that, when expressed
in terms of the parameters $(\alpha,\lambda)$, the mappings turn out to
act as certain $2\times 2$ matrices in the Weyl group of $SU(3)$.
The matrices $\LL$ and $\TT$  defined as
\begin{equation}
\LL=\left(\begin{array}{cc}
1&0\\
0&-1
\end{array}\right)
\quad \, , \quad
\TT=\frac{1}{2}\left(\begin{array}{cc}
-1&\sqrt{3}\\
-\sqrt{3}&-1
\end{array}\right)
\end{equation}
generate the Weyl group of $SU(3)$,  the matrix $\LL$ describing a rotation
and $\TT$ a reflection.  In \cite{DDTb} we found a
 spectral equivalence  between a pair of radial problems
\begin{equation}
\gamma(\bbalpha)D(E,\bbalpha) = \gamma(\TT \bbalpha) D(E,\TT
\bbalpha)~,
\end{equation}
where
$\gamma(\bbalpha)=2\sqrt{i \pi}/ \Gamma (\lambda/ 2\sqrt{3})$.
An equivalence was also obtained at a particular set of points for which
the radial problem $D(E,\bbalpha)$ has a quasi-exactly solvable
sector \cite{Tur,Ush}.  For positive integer $J$ and  $\bbalpha_J =
-(4J+\lambda/\sqrt 3,\lambda)$,
the first $J$ levels of $D(E,\bbalpha_J)$ can be computed exactly as
the zeros of the Bender-Dunne polynomials $P_J(E)$ \cite{BD}.
The result is that modulo the QES levels, the problems $D(E,\TT\LL
\bbalpha_J)$ and
$D(E,\bbalpha_J)$ are isospectral.

A further spectral equivalence particularly relevant to the main theme
of this paper occurs between a radial problem and the related lateral
problem (\ref{PTg}) taken at $M=3$.
The equivalence is
\begin{equation}
T(-E,\bbalpha)=\gamma(\LL\TT\bbalpha) D(E,\LL\TT\bbalpha)~.
\end{equation}
If we denote $ \LL \TT\bbalpha=(\tilde \alpha , \tilde \lambda )^T$
  then the
  problem on the RHS is Hermitian for $\tilde \lambda>0$, and this
relation provides a simple explanation for  the reality of the
spectrum of the $\PT$-symmetric problems in these particular cases.
At the points $\bbalpha=\bbalpha_J$ we can combine the dualities to prove
  a simple result concerning the spectrum of
  $T(E,\bbalpha_J)$, namely that the only energies to become complex
as the parameters $(\alpha,l)$ move into region $A$ through QES values
lie in the solvable part of the spectrum \cite{DDTc}.
 This adds to a similar, as yet unproven, conjecture concerning quartic QES
  potentials \cite{BB4}.

 More detailed
reviews of the ODE/IM correspondence, with more extensive sets of
references, can be found in
\cite{DDTrev,DDTrevb,DDTrevc}.   The original correspondence
relates only the groundstate of a quantum field theory to an ODE,
but recently  a set of differential equations has been found for
the excited states of the field theory \cite{BLZhigh}.
Further aspects of the correspondence are still
being developed, and much remains to be understood.
In particular,  it would
be valuable to have a more physical insight into why the relationship
between integrable
quantum field theories and ordinary differential equations should be
so close.
As more examples are uncovered, we can start to hope that progress on
this so-far mysterious issue may not be too far away.

\bigskip\noindent{\bf Acknowledegments --~}
TCD thanks the organisers for the opportunity to speak at the
conference; TCD and RT thank the UK EPSRC for a Research
Fellowship and an Advanced Fellowship respectively.
This work was partially supported by the EC network ``EUCLID'',
contract number HPRN-CT-2002-00325.

\end {document}